\documentclass{amsart}
\usepackage{amsmath}
\usepackage[utf8]{inputenc}
\usepackage{amssymb}
\usepackage{amsfonts}
\usepackage{float}

\theoremstyle{plain}

\numberwithin{equation}{section}

\Large
\begin{document}
\title[décomposabilité des indicateurs de pauvreté]{Sur la décomposabilité empirique des indicateurs de pauvreté}
\author{$^{1,2,3,*}$Gane Samb Lo}
\author{$^{4}$ Cheikh Mohamed Haidara}

\begin{abstract}
Nous étudions la décomposition empirique des indicateurs de pauvreté. Gr\^{a}ce aux données de l'Enquête Sénégalaise Auprès des Ménages (ESAM) de 1996, nous montrons que le défaut de décomposabilité de ces indicateurs, en particulier ceux de Sen et de Shorrocks,sur la variable revenu, pour plusieurs types de stratification de la population, est pratiquement nul, de l'ordre d'un à deux pour mille. Ceci permet d'utiliser la décomposition des indicateurs de Sen et de Shorrocks sans conséquences fâcheuses. Un modèle explicatif de ces résultats est posé en vue de recherches ultérieures.\\

\noindent (\textbf{English})We study the empirical decomposition of poverty indicators. Given the data from the 1996 Senegalese Households Survey (ESAM), we show that the decomposability gap of these indicators, in particular those of Sen and Shorrocks, on the income variable for several yypes of stratification of the population, is practically zero, of the order of one to two per thousand. This makes it possible to use the decomposition of the Sen and Shorrocks indicators without any untoward consequences. An explanatory model of these results is presented for future research
\bigskip
\bigskip \noindent $^{(4)}$ Cheikh Mohamed Haidara (chheikhh@yahoo.com), Gane Samb Lo (gane-samb.lo@ugb.edu.sn). LERSTAD, Gaston Berger University, Saint-Louis, Senegal.\newline

\noindent $^{(3)}$ LSTA, Pierre and Marie Curie University, Paris VI, France.\newline
\noindent $^{(4)}$ \textit{Gane Samb Lo (gslo@aust.edu.ng)}. AUST - African University of Sciences and Technology,
Abuja, Nigeria\newline

\noindent \textit{$^{\ast}$ Corresponding author}. Gane Samb Lo. Email :
gane-samb.lo@edu.ugb.sn, ganesamblo@ganesamblo.net, gslo@aust.edu.ng.\\
\textit{Permanent address} : 1178 Evanston Dr NW T3P 0J9,
Calgary, Alberta, Canada.

\bigskip
\textbf{NB} : This work was presented at the Conference ATM 2008 (Association Tiers-Monde 2008), held in University Gaston, Saint-Louis, Sénégal. It served as a catalyst of a series of works on the statistical decomposability estimation, starting from the work \cite{haidara-lo}. The paper has been very slightly updated. 
\end{abstract}


\maketitle

\section{Introduction}

L'objectif de cette note est l'étude empirique de la décomposabilit%
é des indicateurs de pauvreté, sur la base des données du Sén%
égal, précisément sur la base de données ESAM I de 1996. La
pauvreté est un phénomène complexe dont l'exploration et
l'analyse requièrent l'utilisation simultanée de plusieurs
disciplines : l'économie, la sociologie, les statistiques, la mé%
decine, etc. Les aspects quantitatifs, à coté des aspects
qualitatifs, jouent un r\^{o}le important. L'appréciation quantitative
se réalise par plusieurs indicateurs de pauvreté dont naturellement
la prévalence de pauvreté, c'est-à-dire la proportion de pauvres
dans la population étudiée.\\

\noindent Ces mesures de pauvreté sont basées sur le revenu ou la dépense
des ménages. Dans la suite, la variable revenu ou dépense $%
Y=(Y_{j},1\leq j\leq N)$\ de la population étudiée de N ménages
dont les valeurs ordonnées sont notées : $Y_{1,N}\leq Y_{2,N}\leq
...\leq Y_{N,N}$. Pour définir un ménage pauvre, on fait appel à
un seuil ou une ligne de pauvreté Z de sorte qu'un ménage j est d%
éclaré pauvre si son revenu est inférieur à Z, c'est-à-dire 
\begin{equation}
(j\text{ pauvre)}\Longleftrightarrow (Y_{j}<Z)  \label{decomp1}
\end{equation}

\bigskip
\noindent Le nombre de ménages pauvres est le nombre Q tel que $Y_{Q,N}<Z\leq
Y_{Q+1,N}$$.$ Les revenus des pauvres sont $\{Y_{1,N},Y_{2,N},...,Y_{Q,N}\}$$%
.$ La détermination du seuil est aussi une question complexe, ayant fait
l'objet d'une multitude de discussions, suivant les approches utilisées
: en termes de conditions de vie minimales, en termes de capacités, en
termes de survie, etc. Le lecteur intéressé peut trouver une bonne
base de discussion dans Ravallion \cite{rav}. Les indicateurs de pauvreté
qui nous préoccupent dans cet article sont des fonctions mathé%
matiques des revenus des Q pauvres sous la forme globale de l'Indicateur de
Pauvreté Général (IPG) introduit dans \cite{lo}, sous la forme 
\begin{equation}
P(N,Y,Q)=\delta (\frac{A(Q,N,Z)}{NB(Q,N)}\overset{Q}{\underset{j=1}{\sum }}%
w(\mu _{1}N+\mu _{2}Q-\mu _{3}j+\mu _{4})\text{ }d\left( \frac{Z-Y_{j,N}}{Z}%
\right) ),  \label{ipg00}
\end{equation}%

\noindent o\`{u}  
\begin{equation*}
B(Q,N)=\sum_{j=1}^{Q}w(j),
\end{equation*}

\noindent o\`u $A\left( \cdot \right) ,$\ $B(\cdot ),$\ $w\left( \cdot \right) ,and$\ $%
d\left( \cdot \right) $\ sont des fonctions mesurables et $\mu _{i}$\ $%
(1\leq i\leq 4$) sont des constantes données. Cet indicateur global
contient, tel que montré dans \cite{lo}, l'essentiel des indicateurs
utilisés dans l'analyse de la pauvreté. Il existe toute une
axiomatique sur ces indicateurs depuis le travail de pionnier de Sen \cite%
{sen}. Il découle de ses travaux et d'autres chercheurs qu'un indicateur
doit posséder certaines propriétés. D'autres propriétés
non obligatoires sont aussi souhaitées. Une revue très complète
de ces concepts se trouve dans \cite{zheng}.\\

\noindent Dans notre terminologie, l'indicateur est pondéré si le poids $%
w(\cdot )$ et la fonction $\delta (\cdot )$ sont des fonctions identiques et 
$A(Q,N,Z)=Q.$ Parmi la classe des indicateurs non pondérés, les plus
connus sont certainement ceux de la classe de Foster, Greer et
Thorbecke(1984) \cite{fgt}, 
\begin{equation}
FGT(Y,N,Q,\alpha )=\frac{1}{N}\overset{Q}{\underset{j=1}{\sum }}\left( \frac{%
Z-Y_{j,N}}{Z}\right) ^{\alpha },\bigskip   \label{decomp3}
\end{equation}

\noindent o\`{u} $\alpha \geq 0$ est un paramètre non négatif. Pour $\alpha =0,
$ cet indicateur se réduit à la proportion de pauvre Q/N; pour $%
\alpha =1,$ il s'appelle intensité de pauvreté et pour $\alpha =2,$
l'aversion contre la pauvreté.\\

\noindent Parmi les indicateurs pondérés, nous retenons les plus célè%
bres. La mesure de Sen (1976) 
\begin{equation}
SE(Y,N,Q)=\frac{2}{N(Q+1)}\sum_{j=1}^{Q}(Q-j+1)\left( \frac{Z-Y_{j,N}}{Z}%
\right)   \label{decomp4}
\end{equation}

\noindent et celle de Shorrocks (1995) 
\begin{equation}
SH(Y,N,Q)=\frac{1}{N^{2}}\sum_{j=1}^{Q}(2N-2j+1)\left( \frac{Z-Y_{j,N}}{Z}%
\right) .  \label{decomp5}
\end{equation}

\noindent Ces deux mesures affectent un poids plus importants pour les plus pauvres.\
Elles sont donc très adaptées pour détecter toute variation sur
les plus pauvres. Nous prendrons cet échantillon de mesures pour notre
banc d'essai d'études empiriques, à coté de la statistique de
Ray(1989) \cite{ran} , dépendant d'un paramètre $\alpha \geq 0:$

\begin{equation}
R(Y,N,Q)=\frac{g}{NZ}\sum_{i=1}^{Q}((Z-Y_{j,N})/g)^{\alpha }
\end{equation}%

\noindent o\`{u} 
\begin{equation}
g=\frac{1}{Q}\sum_{j=1}^{j=Q}(Z-Y_{j,N})
\end{equation}

\noindent est le déficit moyen de pauvreté globale.\\

\noindent Maintenant, nous allons nous focaliser sur une des propriétés des
indicateurs, précisément la décomposabilité. Dans la section
2, qui suit, nous allons la définir et surtout expliquer son importance
et la grande attention que lui attachent tous les théoriciens de la
pauvreté. Enfin dans la section 3, nous prouverons à l'aide de
travaux numériques que tous les indicateurs sont pratiquement dé%
composables, sur les bases ESAM du Sénégal, c'est-à-dire qu'en
les considérant comme décomposables, l'erreur commise est tout à
fait négligeable. 

\section{LA DECOMPOSABILITE}

\subsection{Définition} $ $\\

\noindent Supposons que la population étudiée est divisée en K
sous-groupes identifiables de tailles respectives $N_{1},N_{2},...,N_{K}$\
avec $N_{1}+N_{2}+...+N_{K}=N$, les nombres de pauvres (par rapport au mê%
me seuil Z) respectifs Q$_{1},Q_{2},...,Q_{K}$ avec Q$_{1}+Q_{2}+...+Q_{K}=Q.
$ Soit Y$^{i}=\{Y_{i,1},Y_{i,2},...,Y_{i,N_{i}}\}$ le vecteur revenu du i$^{%
\grave{e}me}$ sous-groupe.\\

\noindent Si nous étudions un indicateur de pauvreté donné dont la valeur
globale sur la population totale est $P(N,Y,Q)$, il peut aussi être é%
valué dans chaque groupe et sa valeur dans le i$^{\grave{e}me}$
sous-groupe est $P(N_{i},Y^{i},Q_{i}).$\\

\noindent L'indicateur est dit décomposable si sa valeur globlale $P(N,Y,Q)$ est
un barycentre de ses valeurs dans les sous-groupes $P(N_{i},Y^{i},Q_{i})$
(pour i=1,2,...,K) affectées des poids positifs $\omega _{i}(N_{i})$ ind%
épendants de la variable Y, c'est-à-dire 
\begin{equation}
P(N,Y,Q)=\sum_{i=1}^{i=K}\omega _{i}\text{ \ }P(N_{i},Y^{i},Q_{i})
\label{decomp6}
\end{equation}%

\noindent avec $\omega _{1}+\omega _{2}+...+\omega _{K}=1.$ Nous excluons de la dé%
composabilité les cas o\`{u} le poids $\omega _{i}$ dépend des
observations Y et aussi de Q, car Q est fonction de celles-ci. La dé%
composabilité souhaitée est celle qui s'opère avec les poids $%
\omega _{i}=N_{i}/N,$ qu'on peut appeler décomposabilité
proportionnelle ou additive. 

\subsection{Importance de cette propriété.} $ $\\

\noindent L'importance de cette propriété est toute pratique. On peut en effet
travailler de manière sectorielle pour lutter contre la pauvreté. Il
sera possible de fixer des catégories cibles, très fragiles et y
combattre la pauvreté. Avec les statistiques officielles, des
informations fiables sur les valeurs des poids peuvent être disponibles.
Dans ce cas, toute réduction de la pauvreté $\Delta
P(N_{i},Y^{i},Q_{i}),$ obtenue dans un sous-groupe, contribue exactement de $%
\omega _{i}\Delta P(N_{i},Y^{i},Q_{i})$ dans la réduction globale. A la
rigueur, des équipes totalement indépendantes peuvent mener les
politiques de réduction de la pauvreté dans ces sous-groupes, ou des
programmes déconnectés peuvent y être implantés. A tout
moment, l'acquis des ces différents programmes peut être ainsi mesur%
é 
\begin{equation}
\Delta P(N,Y,Q)=\sum_{i=1}^{i=K}\omega _{i}\text{ \ }\Delta
P(N_{i},Y^{i},Q_{i})  \label{decomp7}
\end{equation}

\noindent Ceci explique l'importance que les auteurs attachent à cette propriét%
é. En effet, imaginons que les poids dépendent aussi des
observations Y. Sur une petite période au cours de laquelle, on suppose
que les tailles ne bougent pas, nous aurons la formule 
\begin{equation}
\Delta P(N,Y,Q)=\sum_{i=1}^{i=K}\omega _{i}(N_{i},Y^{i})\text{ \ }\Delta
P(N_{i},Y^{i},Q_{i}).  \label{decomp8}
\end{equation}%
\begin{equation*}
+\sum_{i=1}^{i=K}\Delta \omega _{i}(N_{i},Y^{i})\text{ }P(N_{i},Y^{i},Q_{i})
\end{equation*}

\noindent Il faudrait alors une connaissance de la variation $\Delta \omega
_{i}(N_{i},Y^{i})$, ce qui n'est pas toujours possible. Dans tous les cas,
un travail théorique complémentaire est nécessaire. 

\subsection{Décomposabilité des indicateurs usuels.} $ $\\

\noindent Les indicateurs de la classe de Foster-Greer-Thorbecke (FGT) sont dé%
composables. En général, les indicateurs non pondérés de la
forme 
\begin{equation}
P(Y,N,Q)=\frac{1}{N}\underset{j=1}{\sum^{Q}}d\left( \frac{Z-Y_{j,N}}{Z}%
\right)   \label{decomp9}
\end{equation}

\noindent o\`{u} d est une fonction positive croissante, sont décomposables. Par
contre, les indicateurs très importants de Sen et de Shorrocks ne sont
pas décomposable, autrement dit le défaut de décomposabilité 
\begin{equation}
D(Y,N,Q)=P(N,Y,Q)-\sum_{i=1}^{i=K}\omega _{i}\ P(N_{i},Y^{i},Q_{i})
\label{decomp10}
\end{equation}

\noindent est non nul. Ray(1989) obtient une décomposabilité qualifiée de
non additive sous la forme 
\begin{equation}
R(N,Y,Q)=\sum_{i=1}^{i=K}\frac{n_{i}}{N}(\frac{g_{i}}{g})^{\alpha -1}\text{
\ }R(N_{i},Y^{i},Q_{i})  \label{decomp11}
\end{equation}

\noindent dans laquelle 
\begin{equation}
g_{i}=\frac{1}{Q_{i}}\sum_{j=1}^{j=Q_{i}}(Z-Y_{i,j})  \label{decomp12}
\end{equation}

\noindent est le gap moyen de pauvreté dans la $i^{\grave{e}me}$ classe et 
\begin{equation*}
g=\sum_{i=1}^{i=K}\frac{n_{i}}{N}\text{ }(Z-Y_{j})
\end{equation*}

\noindent Ici le poids $\omega _{i}=\frac{n_{i}}{N}(\frac{g_{i}}{g})^{\alpha -1}$ d%
épend de la variable et donc ne rentre pas dans notre étude. Dans ce
dernier cas, la formule (\ref{decomp8}) devient aussi compliquée que 

\begin{eqnarray*}
\Delta R(N,Y,Q)&=&\sum_{i=1}^{i=K}\frac{n_{i}}{N}(\frac{g_{i}}{g})^{\alpha -1}%
\text{ \ }\Delta R(N_{i},Y^{i},Q_{i})\\
&+&\sum_{i=1}^{i=K}(\alpha -1)\frac{n_{i}}{N}(\frac{g\Delta g_{i}-g_{i}\Delta
g)}{g^{2}})(\frac{g_{i}}{g})^{\alpha -2}\text{ \ }R(N_{i},Y^{i},Q_{i})
\end{eqnarray*}

\noindent Il est certain qu'on ne peut évaluer cette formule que par une étude
approfondie. Cependant, nous allons montrer dans la section suivante que le d%
éfaut de décomposabilité est presque nul pour les revenus de la
base ESAM I, pour diverses stratifications.\\

\section{ETUDE EMPIRIQUE DE LA DéCOMPOSABILITé}

Nous utilisons les données de la base de l'Enquête Séné%
galaise Auprès des Ménages (ESAM I) qui a concerné N=3278 mé%
nages. La variable REVTOT est constituée du revenu total du ménage.
La variable EQADUL représente l'équivalence adulte du ménage.
Nous travaillerons avec Y=REVTOT/EQADUL, le revenu annuel individualisé.
Le seuil officiel de 392 francs CFA par jour est utilisé, ce qui donne
Z=392 x 365 francs par année. Nous allons calculer le défaut de d%
écomposabilité : 
\begin{equation}
DDP(Y,N,Q)=P(N,Y,Q)-\sum_{i=1}^{i=K}\omega _{i}\ P(N_{i},Y^{i},Q_{i})
\label{dd01}
\end{equation}

\noindent pour diverses décompositions de la population pour les indicateurs de Sen et de Shorrocks.\\

\noindent Pour chaque cas, nous commençons par donner le nombre de strates $K$ puis les noms des subdivisions.\\

\noindent Les résultats sont probants et concluent que ce défaut est très faible, entre un pour cent et deux pour mille\\

\subsection{Décomposition par rapport aux dix régions} $ $\\

\noindent K=10 sous-groupes : Kolda, Dakar, Ziguinchor, Diourbel, Saint-Louis, Tamba,
Kaolack, Thiès, Louga, Fatick.\\

\noindent \textbf{Estimation du défaut de décomposabilité}.

\begin{center}
\begin{tabular}{|l|l|l|}
\hline
Type de mesures & mesures de Sen & mesure de Shorrocks 
\\ \hline
Mesure décomposée & 0,404 & 0,1706 \\ 
mesure Globale & 0,4066 & 0,1872 \\ 
défaut de décomposabilité & 0,0026 & 0,0165 \\ \hline
\end{tabular}
\end{center}

\bigskip 

\subsection{Décomposition par rapport au genre du chef de ménage} $ $\\

\noindent K=2 sous-groupes : Male et Femelle.\\

\noindent \textbf{Estimation du défaut de décomposabilité }.

\begin{center}
\begin{tabular}{|l|l|l|}
\hline
Type de mesures & mesures de Sen & mesure de Shorrocks \\ \hline
Mesure décomposée & 0,4056 & 0,1857 \\ 
mesure Globale & 0,4066 & 0,1872 \\ 
défaut de décomposabilité & 0,001 & 0,0015 \\ \hline
\end{tabular}
\end{center}

\bigskip 

\subsection{Décomposition par rapport à l'ethnie du chef de mé%
nage}

\noindent K=7 sous-groupes : Wolof, Pular, Serer, Diola, Mandingue, Soninké,
Autres Sénégalais, Africains.\\

\noindent \textbf{Estimation du défaut de décomposabilité}.\\

\begin{center}
\begin{tabular}{|l|l|l|}
\hline
Type de mesures & mesures de Sen & mesure de Shorrocks \\ \hline
Mesure décomposée & 0,4037 & 0,1825 \\ 
mesure Globale & 0,4066 & 0,1872 \\ 
défaut de décomposabilité & 0,0029 & 0,004 \\ \hline
\end{tabular}
\end{center}

\bigskip 

\subsection{Décomposition par rapport à l'état matrimonial du
chef de ménage}

\noindent K=6 sous-groupes : célibataire, marié(e) monogame, marié(e)
polygame, veuf veuve, divorcé(e), Autre.\\

\noindent \textbf{Estimation du défaut de décomposabilité}.

\begin{center}
\begin{tabular}{|l|l|l|}
\hline
Type de mesures & mesures de Sen & mesure de Shorrocks \\ \hline
Mesure décomposée & 0,4056 & 0,1860 \\ 
mesure Globale & 0,4066 & 0,1872 \\ 
défaut de décomposabilité & 0,001 & 0,0011 \\ \hline
\end{tabular}
\end{center}

\bigskip 

\subsection{Décomposition par rapport au niveau d'instruction du chef de
ménage}

\noindent K=5 sous-groupes : Aucun, Primaire, Secondaire, Supérieur, Non Déclaré.\\

\noindent \textbf{Estimation du défaut de décomposabilité}.

\begin{center}
\begin{tabular}{|l|l|l|}
\hline
Type de mesures & mesures de Sen & mesure de Shorrocks \\ \hline
Mesure décomposée & 0,4058 & 0,1690 \\ 
mesure Globale & 0,4066 & 0,1872 \\ 
défaut de décomposabilité & 0,0008 & 0,0182 \\ \hline
\end{tabular}
\end{center}

\subsection{Conclusion} $ $\\

\noindent Nous constatons la décomposabilité pratique des indicateurs
importants de Sen et Shorrocks. Comment expliquer une telle performance?
Nous tentons une première explication ici-bas. 

\section{ESTIMATION DU DEFAUT DE DECOMPOSABILITE.}

Reprenons la forme globale de l'IPG, 
\begin{equation}
P(N,Y,Z)=\delta (\frac{A(Q,N,Z)}{NB(Q,N)}\overset{Q}{\underset{j=1}{\sum }}%
w(\mu _{1}N+\mu _{2}Q-\mu _{3}j+\mu _{4})\text{ }d\left( \frac{Z-Y_{j,N}}{Z}%
\right) ).  \label{gpi01}
\end{equation}

\noindent Le défaut de décomposabilité se note.

\begin{equation*}
DDP(N,Y,Q)=P(N,Y,Q)-\sum_{i=1}^{k}W_{i}\text{ }P(N_{i},Y^{i},Q_{i}).
\end{equation*}

\noindent Si nous sommes en mesure d'estimer la quantité DDP par intervalle de
confiance à un seuil acceptable, il nous sera loisible d'étudier
l'indice séparément dans chaque strate et de revenir à un niveau
global. L'étude empirique précédente a conclu que ce nombre est
presque nul, ce qui aurait correspondu dans notre cas, à un intervalle
de confiance très serré sur zéro.\\

\noindent Les résultats de (\cite{sslo}) permettent une première explication.
En effet, l'indice de pauvreté $P(N,Y,Z)$ est estimé gr\^{a}ce à
un sondage et sa valeur échantillonnée est 
\begin{equation*}
p(n,G,Z)=\delta (\frac{A(q,n,Z)}{n\text{ }B(q,n)}\overset{q}{\underset{j=1}{%
\sum }}w(\mu _{1}n+\mu _{2}q-\mu _{3}j+\mu _{4})\text{ }d\left( \frac{%
Z-Y_{j,n}}{Z}\right) ),
\end{equation*}

\noindent o\`{u} $q$ est le nombre de pauvres dans l'échantillon et $Y_{1,n}\leq
...\leq Y_{n,n}$ sont les revenus échantillonnés ordonnés. F est
la fonction de répartition commune du revenu au plan global. On peut
fixer la taille de l'échantillon à n, puis tirer $n_{i}$ individus
dans le i$^{\grave{e}me}$\ sous-groupes et obtenir l'indice échantilonn%
é correspondant au i$^{i\grave{e}me}$ sous-groupe, 

\begin{equation*}
p(n_{i},G_{i},Z)=\delta (\frac{A(q_{i},n_{i},Z)}{n_{i}\text{ }B(q_{i},n_{i})}%
\overset{q_{i}}{\underset{j=1}{\sum }}w(\mu _{1}n_{i}+\mu _{2}q_{i}-\mu
_{3}j_{i}+\mu _{4})\text{ }d\left( \frac{Z-Y_{j,n_{i}}^{i}}{Z}\right) )
\end{equation*}

\noindent o\`{u} $q_{i}$ est le nombre de pauvres dans le i$^{\grave{e}me}$ é%
chantillon et $Y_{1,n_{i}}^{i}\leq ...\leq Y_{n_{i},n_{i}}^{i}$ sont les
revenus ordonnés dans ce sous groupe et $F_{i}$\ est la fonction de r%
épartition du revenu dans le i$^{\grave{e}me}$\ sous-groupe. De plus, 
\begin{equation*}
n_{1}+n_{2}+...+n_{k}=n\text{ \ and \ }q_{1}+q_{2}+...+q_{k}=q.
\end{equation*}

\noindent Alors 
\begin{equation}
dd_{n}=p(n,G,Z)-\sum_{i=1}^{k}W_{i}\text{ }p(n_{i},G_{i},Z)  \label{dd02}
\end{equation}

\noindent est le défaut de décomposabilité echantillonné. Introduisons
quelques notations pour pouvoir utiliser les travaux antérieurs. Posons 
\begin{equation*}
y_{0}(i)=\inf \{x,G_{i}(x)>0\},
\end{equation*}

\noindent et par la suite,
\begin{equation*}
X^{i}=1/(Y^{i}-y_{0}(i)),
\end{equation*}

\noindent de fonctions de répartition 
\begin{equation*}
F_{i}(\cdot )=1-G_{i}(1-1/\cdot ).
\end{equation*}

\noindent Sous certaines conditions assez douces, précisées dans \cite{sslo}
, il est démontré que 
\begin{equation*}
p(n_{i},G_{i},Z)\rightarrow _{p}\delta (D_{i})=\delta
(\int_{0}^{G_{i}(Z)}L(s)\text{ }d\left( \frac{Z-y_{0}(i)-1/F_{i}^{-1}(1-s)}{Z%
}\right) ds),
\end{equation*}

\noindent o\`{u} $L(\cdot)$ est une fonction poids ne dépendant que de la forme
de l'indice et non du sous-groupe. L'absence d'indice dans cette formule
concerne la population globale. En réalité, (\cite{sslo}) donne la
normalité asymptotique mais nous sommes concernés ici que par la
consistance. Il s'en suit alors que 
\begin{equation}
dd_{n}\rightarrow DDP(F,F_{1},...,F_{k})={\large \delta (D)-}%
\sum_{i=1}^{k}W_{i}\text{ }\delta (D_{i}).  \label{dd03}
\end{equation}

\noindent Nous interprétons que dans une population de taille très grande, la
quantité (\ref{dd01}) calculée à partir de la base est la valeur 
échantillonnée du \ défaut théorique de décomposabilité 
\begin{equation}
DDP(F,F_{1},...,F_{k})={\large \delta (D)-}\sum_{i=1}^{k}W_{i}\text{ }\delta
(D_{i}).  \label{dd04}
\end{equation}

\noindent Il joue alors le r\^{o}le de (\ref{dd02}) et devient de ce fait une
approximation de (\ref{dd04}). La formule (\ref{dd03}) permet alors de
conclure que (\ref{dd01}) est une estimation asymptotiquement sans biais de (\ref{dd04}). On peut vérifier facilement que lorsque le poids est
unitaire, cette estimation est sans biais à distance finie.\\

\bigskip \noindent \bigskip Ce résultat permet d'expliquer les résultats des
simulations. En effet lorsque la distribution du revenu est égale sur la
population, le défaut de décomposabilité est asymptotiquement
nul pour tous les indicateurs de la classe IPG puisque dans ce cas, 

\begin{equation*}
{\large \delta (D)=}\delta (D_{1})=\delta (D_{2})=...=\delta (D_{k}).
\end{equation*}

\noindent De même si les distributions sont sensiblement les mêmes et ne s'%
écartent pas trop les unes des autres, la distribution globale F reste
sensiblement proche et le défaut de décomposabilité reste trè%
s faible à l'infini. Il en est ainsi dans le cas du Sénégal. En
effet, le modèle lognormal est globalement accepté pour les
distributions F$_{i}$ des revenus ou dépenses. Les valeurs des paramè%
tres $\sigma _{i}$\ et $m_{i}$ estimés tournent globalement autour de $%
\sigma =1$ et $m=-12$. Ceci est une explication plausible des résultats
de nos simulations. 

\section{Conclusion}

Sur la base des données ESAM I du Sénégal et pour diverses
stratifications, le défaut de décomposition s'élève en général de un à deux pour mille, ce qui est négligeable pour les
ordres de grandeurs mesurées de l'ordre moyen de 30\%. Pour deux cas, le
défaut s'élève à un pour cent, ce qui est encore largement
acceptable. Notre étude conclut sur le fait que les mesures de sen et de
Shorrocks sont pratiquement décomposables, avec une erreur de un pour
cent à deux pour mille. En conséquence, recommandation est faite aux
acteurs de la lutte contre la pauvreté, d'utiliser la décomposabilit%
é empirique des mesures de Sen et de Shorrocks.\\ 

\noindent Les travaux de \cite{lo}, explique aussi cela pour une distribution faiblement variable sur les sous-groupes. Il en est ainsi lorsque le modèle lognormal est accepté avec une faible variation des paramètres des lois lognormales, ce qui est le cas dans
les bases ESAM. Néanmoins, il reste à estimer par intervalle de
confiance le défaut de décomposabilité, nul ou non. Cette
estimation permettra de décharger leschercheurs et utilisateurs du fardeau de la contrainte de favoriser les indices décomposables au risque de perdre les avantages de
ceux qui ne le sont pas, puisqu'elle permettra de recomposer l'indice
global, après une étude sectorielle.\ Des recherches dans ce sens
sont en cours.\\

\bigskip \noindent \textbf{Acknowledgement}. Ont particip\'e aux \'etude de simulation : Mamadou A. Niang, Bassirou Diagne, Papa A. Camara, Steve A. Matongo, Mohamed S. Ould Mouhamed.


\begin{thebibliography}{9}
\bibitem{fgt}  Foster, J.E, J. Greer, and Thorbecke(184). A class of
decomposable Poverty Measure Indices. \textit{Econometrica}, 52, 761-766.

\bibitem{haidara-lo} Haidara M.C and LO G.S.(2012) Statistical Estimation of
Gap of Decomposability of the General Poverty Index. \textit{International
Journal of Statistics and Probability}, 1 (2). doi:10.5539/ijsp.v1n2p211

\bibitem{sslo}  LO, G. S., Sall, S.T. and seck, C. T. (2007). The
General Asymptotic Theory of Poverty Measures. Submitted. 

\bibitem{lo}  LO, Gane Samb (2013). On the General Poverty Index. Far East Journal of Theoretical Statistics. Volume 42. (1), 1-22. 

\bibitem{rav}  Martin Ravallion (1992). Poverty Comparisons. A Guide
to Concepts and Methods. Lsms, Working Paper, 88, WorldBank.

\bibitem{ran}  Ranjan Ray(1989), A new class of decomposable poverty
measures. \textit{Indian Economy Journal}, vol.36, 30-38. 

\bibitem{sen}  Sen Amartya K.(1976). Poverty : An Ordinal Approach
to Measurement. \textit{Econometrica}, 44, 219-231.  

\bibitem{taleb}  Taleb Ely Ould Taleb Ahmed(2003), \textit{Les
mesures de la pauvret\'{e}, de l'in\'{e}galit\'{e} et l'impact de la
croissance \'{e}conomique}, m\'{e}moire de DEA, Universit\'{e} Gaston Berger
de Saint-Louis.  

\bibitem{zheng}  Zheng, B.(1997). Aggregate Poverty Measures.
Journal of Economic Surveys, 11 (2), 123-162. 
\end{thebibliography}
\end{document}